\documentstyle[pre,aps,epsf,multicol]{revtex}
\begin{document}

\title{Critical branching-annihilating random walk of two species}
\author{G\'eza \'Odor}
\address{Research Institute for Technical Physics and Materials Science, \\
H-1525 Budapest, P.O.Box 49, Hungary}    
\maketitle

\begin{abstract}
The effect of blocking between different species occurring in one 
dimension is investigated here numerically in the case of particles following 
branching and annihilating random walk with two offsprings. It is shown that 
in two dimension simulations confirm the field theoretical results with 
logarithmic corrections. In one dimension however if particles exhibit 
hard core interaction I confirm the very recent predictions of Kwon {\it et al.} 
[PRL {\bf 85}, 1682 (2000)] that there are two different universality classes 
depending on the spatial symmetry of the offspring production characterized 
by $\beta_S=0.5$ and $\beta_A=2$.
Elaborate analysis of simulation data shows that the order parameter exponent 
$\beta$ does not depend on initial conditions or on diffusion rates of species
but strong correction to scaling is observed. 
By systematic numerical simulations the critical point properties have 
been explored and initial condition dependence of the dynamical exponents 
$Z$ and $\alpha$ is shown. In the case of a random initial state the 
particle-density decay at the critical point follows the $t^{-1/4}$ law
with logarithmic corrections.
\end{abstract}
\pacs{\noindent PACS numbers: 05.70.Ln, 82.20.Wt}

\begin{multicols}{2}
%=============================================================
\section{Introduction}
%=============================================================

Branching and annihilating random walk (BARW) process one of the simplest 
prototypes of a particle system exhibiting nonequilibrium phase transitions.
For small creation/annihilation ratio the system evolves to a completely empty
state from which no return is possible. This so called "absorbing state" lacks
any fluctuations. For higher creation/annihilation rates, a steady state with some
finite concentration of particles emerges with a phase transition.
This phase transition has been investigated first by numerical simulations
\cite{Gras84,Taka,Jensen} and later with field theoretical methods 
\cite{Cardy-Tauber}.
In general, we define annihilation of $k$ particles $k A\to\emptyset$ 
with the rate $\lambda$ and generation of $m$ offsprings $A \to (m+1) A$ 
with rate $\sigma$.

In $D>2$ dimensions the phase transition of these systems can 
be described by mean-field theory but in low dimensions the 
effect of fluctuations introduces more complex behavior. 
The $k=1,2$; $m=1$ case is the so-called directed percolation 
process; the continuous phase transition of it defines a robust universality 
class. According to the hypothesis of Janssen and Grassberger
\cite{Jan81,Gras82} all continuous phase 
transitions to a single absorbing state in homogeneous 
systems with short-ranged interactions belong to this class
provided there is no additional symmetry and quenched randomness present.
As it was shown by Cardy and Ta\"uber \cite{Cardy-Tauber} 
this is the generic class for all critical models with
$k=1,2$; $m=$ odd and indeed a large number of model systems
\cite{Dick-Mar} have been found to belong to this class. 

If $k=m=2$ the parity of the number of particles is conserved locally
under the reactions and another universality class appears:
the so-called parity-conserving (PC) or BARW2 class. 
Particles following BARW2 dynamics may also appear as kinks between 
ordered domains in systems exhibiting two absorbing states  
\cite{Gras84,Men94,Park94,Hin97,Bassler}.
For recent reviews see \cite{Hin2000,MeOdof}.
For $k>2$ the field theory does not predict any new universality classes.

Our knowledge of multiparticle systems is quite incomplete yet. 
A very recent paper of Janssen \cite{Janssen-col} shows that this 
usually leads to asymmetries between pairs of species and 
unidirectionally coupled systems arise. 
These systems have been shown to describe certain surface roughening 
processes \cite{uni-DP,uni-PC}, for example.

The bosonic field theoretical work of \cite{Cardy-Tauber} 
predicts a new universality class for BARW2 process of
symmetric $N$ species (N-BARW2), exhibiting parity conservation 
($m=k=2$) of particles.
They found that offspring creations of the same type 
like $A\to AAA$ are not relevant processes because 
new particles are annihilated immediately.
In the case of branching processes like $A\to ABB$ 
and $A\to BBA$, all $N>1$ component models have the same Feynman 
diagrams as that of the $N\to\infty$ model
and the critical point for $D\le 2$ happens 
to be at $\sigma=0$ zero branching rate
with the following critical exponents:
\begin{equation}
\nu_{||}=2/d,\quad\nu_{\perp}=1/d,\quad Z=2,\quad\alpha=d/2,\quad\beta=1.
\label{FTexps}
\end{equation}
The critical exponents are defined by the usual way. 
The singular behavior of the spatial and temporal correlation 
lengths are characterized by 

\begin{equation}
\xi \propto \sigma^{-\nu_{\perp}} \ \ ,
\end{equation}
\begin{equation}
\tau \propto \sigma^{-\nu_{||}} \ \ .
\end{equation}
The anisotropy or dynamical exponent is $Z = \nu_{||} / \nu_{\perp}$,
while the particle density near the critical point obeys the
scaling law
\begin{equation}
\rho (t,\sigma) \propto t^{-\alpha} \phi( t \sigma^{\nu_{||}}) \ \ ,
\label{rhoscal}
\end{equation}
such that in the $t\to\infty$ steady state limit the order parameter
scales as $\rho \propto \sigma^{\beta}$; hence 
\begin{equation}
\beta = \alpha \nu_{||} = \alpha Z \nu_{\perp}
\end{equation}
scaling relation holds.
At the critical dimension ($D_c=2$), 
logarithmic corrections to density decay are expected.
For $D>2$ the critical point is also at $\sigma=0$
but governed by mean-field exponents \cite{Cardy-Tauber}.

In this work I investigated, numerically, the critical
behavior of the 2-BARW2 model in one and two dimensions
with the motivation that the existing field theories 
of reaction-diffusion models deal with a coarse grained, 
bosonic description of the particle system and therefore 
particle exclusion has not been taken into account. 
In a recent paper \cite{gdkcikk} we have shown that 
in the case of one-dimensional (1D) annihilating random walk 
of two species, the blocking between different species 
can cause marginal perturbations. 
Therefore it is an open question whether 
this can also be observed if we add particle creation
to this process.

In a more recent paper Kwon {\it et al.} \cite{Kwon} argue
based on local mean-field approximation that in this 
case new universality classes emerge indeed.
I confirm the predictions of \cite{Kwon}
for the order parameter exponent by precise numerical 
simulations and show that this exponent does not depend 
on initial-conditions or diffusion ratios of species.
Furthermore I investigate the very nontrivial
time dependence of this model and show numerical evidence
for the initial condition dependence of the dynamical 
exponents and logarithmic corrections.

%=========================================================================
\section{Definition of the model}
%=========================================================================

A random sequential processes have been implemented in $D=1$ and $D=2$ 
dimensional lattices possessing periodic boundary conditions with 
one time step described as follows:

(1) A particle is selected randomly

(2) A nearest neighbor site is selected randomly

(3) With probability $1-\sigma$ the particle is attempted to 

\ \ \ \ \ move to the new site if that is empty

(4) If the new site is filled with a same kind of particle, 

\ \ \ \ \ both of them are removed

(5) With probability $\sigma$ two off-springs of different types

\ \ \ \ \ from the ancestor are placed at two free nearest 

\ \ \ \ \ neighbor sites selected randomly.

In this way I investigate the same model as \cite{Kwon}.
The field theoretical model of \cite{Cardy-Tauber} also
deals with the same particle reactions albeit without exclusion.

%=========================================================================
\section{Simulations from random initial state in two dimensions}
%=========================================================================

The initial conditions for systems with linear sizes $L$ 
have been set up in such a way that $L^2/2$ randomly 
selected A or B particles are placed at random sites.

\subsection{Zero branching rate case ($\sigma=0$)}

\begin{figure}
\epsfxsize=70mm
\centerline{\epsffile{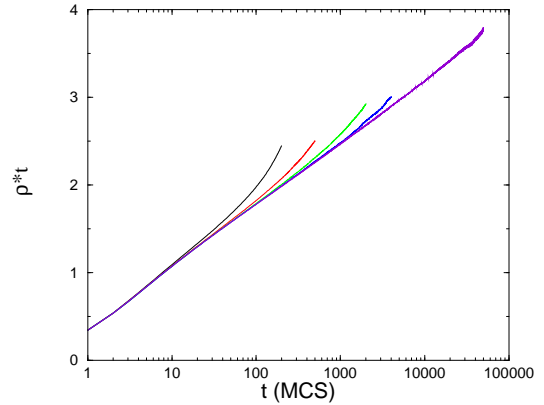}}
\vspace{2mm}
\caption{Density decay (in $2D$) of the total particle number for 
different system sizes: $L=16, 32, 64, 128, 512$ (from top to bottom). 
Throughout the paper $t$ is measured in units of 
Monte Carlo sweeps (MCS) of the lattices.
\label{decayu}
}
\end{figure}

First the field theoretical expectations \cite{Lee,Cardy-Tauber}
for the density decay ($\rho\propto \ln(t) / t$)
were tested. To see the logarithmic corrections, I plotted
$\rho(t)\times t$ versus $\ln(t)$.
As Fig.~\ref{decayu} shows the curves corresponding to
different sizes show linear behavior for short times. 
Later the effect of finite system sizes causes this 
behavior to break down. 
The crossover happens for larger and larger 
times as $L$ goes to $\infty$. 

For the largest system investigated here ($L=512$) a 
logarithmic fitting was applied in the 
$10 < t < 10^4$ region, resulting in
\begin{equation}
\rho(t)= (0.442(2)+0.296(1)\ln(t))/t
\end{equation}
Hence the field theoretical prediction $\alpha=1$ 
with logarithmic corrections has been confirmed by these 
simulations. The amplitude of the logarithmic term however 
differs from that of the pure annihilation model: 
$A_2=1/8\pi$ \cite{Lee}. One may speculate that in the 
two-component system this value is exactly $1/\pi$.

%=============================================================
\subsection{Off-critical case ($\sigma>0$)}
%=============================================================

Traditional, direct simulations of off-critical exponents 
usually produce less precise estimates than 
cluster simulations for example,
but now I could eliminate the two most important 
drawbacks of this method.
The critical point is known to be exactly at $\sigma=0$;
therefore this does not cause uncertainty in
the estimation of the order parameter exponent.
Furthermore I use precise analysis of the data by 
calculating local slopes that enables us to follow 
corrections to scaling.

\begin{figure}
\epsfxsize=70mm
\centerline{\epsffile{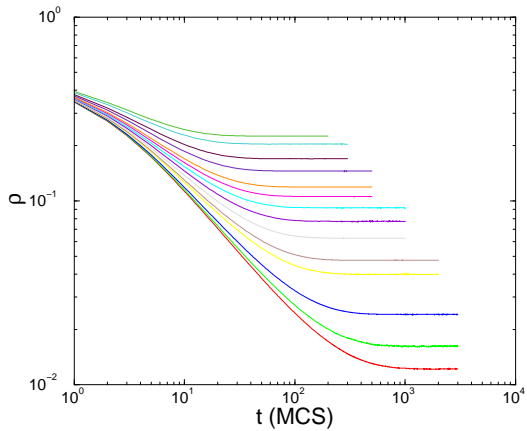 }}
\vspace{2mm}
\caption{Density decay (in $2D$) of the total particle number for 
$\sigma=0.0075,0.01 ... 0.17$ (from bottom to top) in a system of size
$L=512$.
\label{rho_512u}
}
\end{figure}

\begin{figure}
\epsfxsize=70mm
\centerline{\epsffile{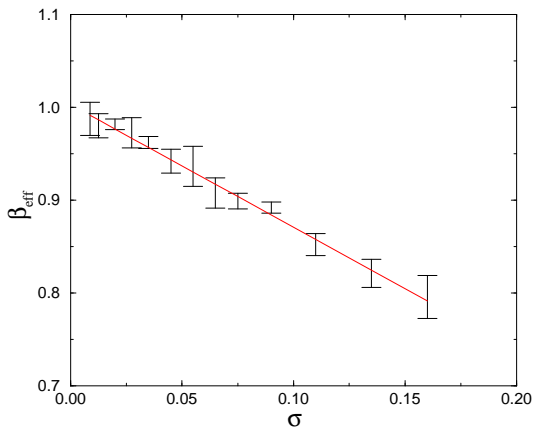}}
\caption{Effective order parameter exponent ($\beta$) results in
$2D$ systems of linear sizes $L=512$.
\label{betaeffu}
}
\end{figure}

The simulations with branching were performed in 
systems of linear sizes: $L=512$ in the 
neighborhood of the critical point 
($\sigma_c=0$) at $\sigma=0.0075,0.01 ... 0.17$ (Fig.~\ref{rho_512u}).
About $100$ samples were run from independent random initial 
states and averaged over for each $\sigma$.
From the results at the critical point (Fig.~\ref{decayu})
one can see that this system size is large enough 
for finite size corrections to be negligible for
$t < 10^4$ MCS investigated here.
That means we can assume that in my steady state
simulations the correlation length does not exceed $L$.
The particle density curves were averaged following 
saturation and the effective $\beta_{eff}(\sigma)$ 
exponents were calculated as
\begin{equation}
\beta_{eff} (\sigma) = \frac {\ln \rho_{i} -\ln \rho_{i-1}}
              {\ln \sigma_i - \ln \sigma_{i-1}} \ \ ,
\end{equation}
providing an estimate for the true asymptotic behavior of the
order parameter
\begin{equation}
\beta = \lim_{\sigma\to 0} \beta_{eff}(\sigma) \,.
\end{equation}
By applying a linear extrapolation as in Fig.~\ref{betaeffu} 
a very good agreement with the field theoretical prediction for 
the order parameter exponent $\beta=1.003(3)$ can be found.

%=========================================================================
\section{Simulations in one dimension}
%=========================================================================

%=============================================================
\subsection{The zero branching rate case ($\sigma=0$) \label{sigmac}}
%=============================================================

In \cite{gdkcikk} we showed that in the case of special 
initial conditions
\begin{verbatim}
....A..A...B.B...B..B.A.....A......B..B...
\end{verbatim}
that assures that the time evolution never freezes 
(i.e. every particle has a neighboring particle of the same 
type to annihilate with) 
the density decays non-universally in an initial-condition 
dependent way.
In this case particles can be regarded as boundaries of 
compact domains of different species that cannot overlap 
each other. This happens in the genaralized Domany-Kinzel 
cellular automaton, for example, \cite{hayeGDK}.

The decay depends on the initial density 
(and hence on the initial interparticle 
separation length) of the species.
We argued on the basis of symmetry considerations 
and by using an analogy to the works of Kaiser and Turban 
\cite{Kaiser-Turban} that this decay follows the powerlaw 
with continuously changing exponents.
We also showed that in the case of asymmetric initial 
conditions, where a small seed of B particles is inserted 
in the see of A's, the cluster-survival probability of
B's (characterized by the exponent $\delta$) depends 
continuously on the initial density of A's 
similar to the case of the pair contact 
process \cite{PCP}. Therefore in this work I 
investigated whether the static exponents show 
any initial-condition dependence as well 
and the form of the scaling law that connects them
to the dynamic ones.

In the case of random initial conditions particles segregate into
coarsening mosaic of alternating domains containing alternating
sequences of A's and B's. By mapping this onto diffusion-controlled 
two-species annihilation ($A+B\to\emptyset$) \cite{DC2A}
Krapivsky and Redner argue that the particle density decays as 
$t^{-1/4}$ \cite{Krap-Red}.

\begin{figure}
\epsfxsize=70mm
\centerline{\epsffile{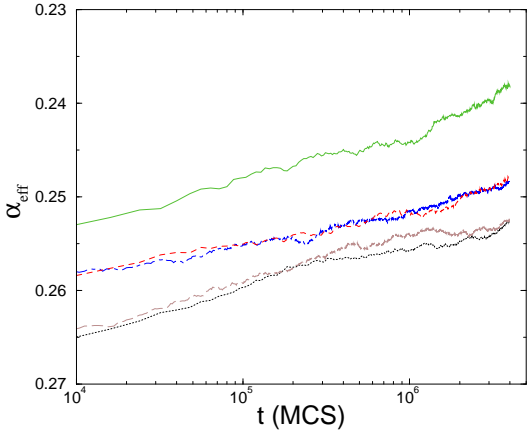}}
\caption{Effective exponent of the particle decay in the
one dimensional BARW2 with two species. 
The initial condition is randomly distributed A's 
and B's along the system of size $L=10^5$ sites. 
The solid line corresponds to initial densities:
$\rho_A=\rho_B=0.375$, 
the dashed line to:
$\rho_A=\rho_B=0.25$, 
the dotted line to
$\rho_A=\rho_B=0.125$.
The dot-dashed line corresponds to $L=10^4$ and
$\rho_A=\rho_B=0.25$,
while the long-dashed line corresponds to $L=10^4$ and
$\rho_A=\rho_B=0.125$
\label{alpha}
}
\end{figure}

I tested this by simulating the process on $L=10^5$ lattices
up to $t=4\times10^6$ MCS. The local-slopes curves 
of the density decay defined as
\begin{equation}
\alpha_{eff}(t) = {- \ln \left[ \rho(t) / \rho(t/m) \right] 
\over \ln(m)} \label{slopes}
\end{equation}
(where usually I use $m=8$) converge to $\sim 0.25$ indeed
(Fig. \ref{alpha}), but again like in \cite{gdkcikk} 
dependence on the initial conditions can be observed.
Now however, the local slopes do not saturate in the
$t\to\infty$ limit as in the case of "pairwise" initial 
conditions \cite{gdkcikk} but show logarithmically 
decreasing behavior as $t\to\infty$. 
This is not an artifact of the finite system 
sizes since the same behavior can be seen on much 
smaller ($L=10^4$) lattices too.
The origin of the logarithmic correction to scaling
is not clear, but it is likely that marginal
perturbation of species on each other causes it. 
Note that logarithmic correction to time dependence 
of the interparticle distances in case of the 
$A+B\to\emptyset$ model have also been shown by
simulations \cite{Ley-Red} and explained to be the
consequence of density inhomogeneity inside the
domains.
This anomalous decay behavior cannot be described
by the conventional field theoretical description of 
a particle system \cite{Cardy-Tauber} in $D=1$ that
omits site restrictions and predicts $\alpha=1/2$.
Also the paper of Kwon {\it et al.} \cite{Kwon} based on
local mean-field approximation does not show 
any initial-condition dependency but predicts
simply the $\alpha=1/2$ result.

\begin{figure}
\epsfxsize=70mm
\centerline{\epsffile{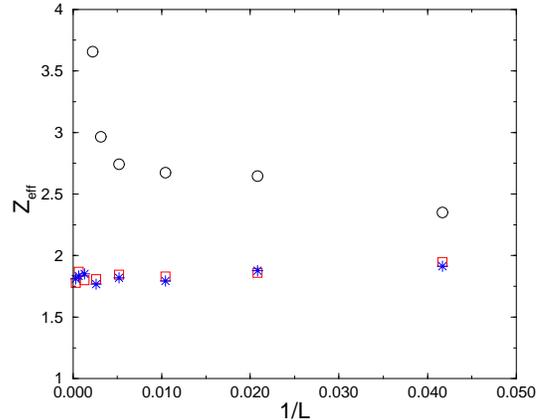}}
\caption{$Z_{eff}$ vs $1/L$ at $\sigma=0$.
The square and star symbols correspond to pairwise initial
conditions with $\rho_A=\rho_B=0.125, 0.25$ respectively.
The circles correspond to random initial conditions with 
$\rho_A=\rho_B=0.0625$.
\label{tausl}
}
\end{figure}

The dynamical exponent $Z$ has been determined through
the measurement of the characteristic time $\tau(L)$
at the critical point. In this study I define $\tau(L)$
as the time it takes for some portion of the sample to
enter the absorbing state. In the case of pairwise initial
conditions this portion was 50\% but in case of random
initial conditions, where the evolution is very slow in
finite systems, I chose this portion to be 10\% in order to
get results for $L>1000$ sizes. We expect the following
finite size scaling at $\sigma=0$
\begin{equation}
\tau(L) \propto L^Z \ \ .
\end{equation}
The data points have been determined within the 
$L=16-4096$ size region and to make precise 
extrapolation, local-slopes analysis has been performed.
In Fig.\ref{tausl} I have plotted 
\begin{equation}
Z_{eff} (L) = \frac {\ln \tau_{i} -\ln \tau_{i-1}}
              {\ln L_i - \ln L_{i-1}} \ \ ,
\end{equation}
as a function of $1/L$. 
As one can see, in the pairwise initial condition
case $Z_{eff}$ converges to $Z=1.82(2)$ as $1/L\to 0$
quite smoothly.
However in the random initial condition case a huge
correction to scaling can be observed and $Z_{eff}$
seems to extrapolate to $Z \sim 4.0(2)$, which is in
accordance with the $t^{-1/4}$ density decay law.
The $Z=1.82(2)$ value is in agreement with our previous
finding \cite{gdkcikk} where we showed that in the case of
pairwise initial conditions the density decays as 
$\sim t^{-0.545(10)}$. We argued there that the 
interparticle distance of AA and BB pairs 
$l_{AA}=l_{BB} \propto \rho(t)^{-1}$ increases with the same
power-law as the region of confinement that can be 
characterized with the exponent $1/Z$. 
Now I could not obtain significant initial-condition 
dependence in $Z$ owing to the lack of such high 
precision simulations that we achieved in \cite{gdkcikk}.
Note that the usual scaling relation
between the single cluster spreading exponent $z$
(describing the mean distance from the origin
$R(t)\propto t^{z/2}$) and the
dynamical exponent $Z=2/z$ is violated. This may be
understood if we realize that in the seed growing
process there is no symmetric, mutual exclusion 
between different species therefore the exponent
$z$ cannot describe the multisurface effect
that arises in the case of pairwise initial conditions.
Neither of these $Z$ values is in agreement
with the assumptions of \cite{Kwon} and the field
theoretical results for a 1D system without particle
exclusion \cite{Cardy-Tauber}.

%=============================================================
\subsection{The off-critical case ($\sigma>0$) 
with symmetric (static) branching \label{static}}
%=============================================================

The effect of the spatial symmetry in simple one-dimensional
BARW processes was first investigated in \cite{Kw-P} and it 
was shown that reflection-symmetric (static) branching causes 
reentrant phase diagrams unlike the asymmetric (dynamic) branching
process.
First with the same special, pairwise initial conditions as 
in our recent paper \cite{gdkcikk}, I performed steady state 
simulations such that offsprings are generated symmetrically
on the left and the right sides of parents 
$A \to BAB$, $B \to ABA$ (static branching).
Kwon {\it et al.} \cite{Kwon} reported
different critical behavior than what was expected
by applying the results of the bosonic field theory 
\cite{Cardy-Tauber} for this particle system.

\begin{figure}
\epsfxsize=70mm
\centerline{\epsffile{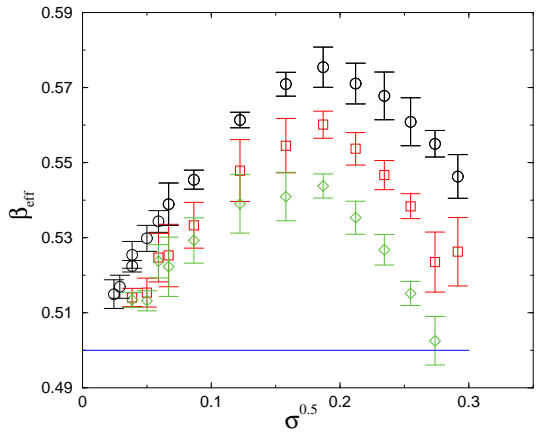}}
\vspace{2mm}
\caption{Effective order parameter exponent ($\beta$) results
for the symmetric branching case.
Circles correspond to the equal diffusion case, squares and 
triangles to $d_A=d_B/2$ nonequal diffusion rate
case of species.
\label{betas}
}
\end{figure}

Similar to the two-dimensional case, the density decay was 
followed carefully and averaging was done within a long time 
period, following the steady state having been built up.
Figure \ref{betas} shows my simulation results 
of the order parameter $\beta$ in cases of 
different diffusion rates. 
As one can see, all the curves show strong 
(nonmonotonic) correction to scaling in the 
neighborhood of $\sigma=0$ but they all
tend to the value $\beta=0.5$. The overshooting of
the effective exponent signals possible logarithmic 
correction to scaling. The relaxation times to steady 
state were $t<10^5$ MCS in these simulations.

I found that in the neighborhood of the critical
point the steady state concentration is unchanged if 
I use larger lattice sizes 
($L=4\times10^5$ instead of $10^5$), 
therefore the strange shape of the curves cannot be 
an artifact of the finite system size of the simulations.
Note, that had we done a simple fitting onto the data
points we could have estimated $\beta\sim 0.55$ 
instead of $\beta=0.5$.
The results are plotted as functions of 
$\sigma^{0.5}$ determined experimentally such that 
the curves approach $\sigma=0$ in an
asymptotically straight line manner. 
This means that the leading correction to the scaling
exponent is $\sim0.5$.
As one can see, the curves converge to $\beta=0.50(1)$ 
in agreement with the predictions of
\cite{Kwon}.
The simulations with other initial densities and
random initial distributions of particles resulted 
in the same steady state values and hence $\beta$ is 
unchanged.

Comparing the $Z$'s of the previous section 
with the $\nu_{\perp}=1/2$ of \cite{Kwon} determined 
by static finite size scaling for the symmetric branching
model we predict for pairwise initial conditions 
$\nu_{||}^p = Z \nu_{\perp} = 0.915(2)$ and
$\nu_{||}^r= 2.0(1)$ for random initial conditions.

The scaling law of density is fulfilled within error limits
$\alpha^p\simeq 0.545\simeq\beta / \nu_{||}^p = 0.5 / 0.915$
for the pairwise initial case,
$\alpha^r\simeq 0.25 \simeq \beta / \nu_{||}^r = 0.5 / 2$
for random initial conditions.

\begin{figure}
\epsfxsize=70mm
\centerline{\epsffile{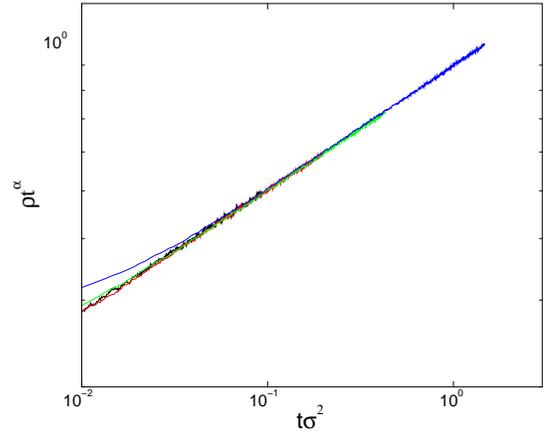}}
\vspace{2mm}
\caption{Data collapse for off-critical simulations of the
symmetric branching model according to the scaling form
(\ref{rhoscal}) with $\alpha^r=1/4$ and $\nu^r{||}=2$.
The different curves correspond to 
$\sigma=0.0005, 0.001, 0.002, 0.005$ and
random initial conditions.
\label{srcol}
}
\end{figure}

The scaling form (\ref{rhoscal}) can be checked by plotting 
$\rho t^{\alpha}$ against 
$t\sigma^{\nu_{||}}$. As Fig. \ref{srcol} shows, one
can obtain a fairly good data collapse with $\nu^r_{||}=2$ and
$\alpha^r=0.25$ (corresponding to random initial conditions)
for large times. For small times the collapse is less good
owing to the effect of logarithmic corrections to $\alpha^r$.
For pairwise initial conditions a very good data collapse
can be achieved with the standard scaling form (\ref{rhoscal})
using $\alpha^p=0.545$, $\nu_{||}^p=0.915$

%=============================================================
\subsection{The off-critical case ($\sigma>0$) with asymmetric 
(dynamical) branching}
%=============================================================

For the same special, pairwise initial conditions as in 
our recent paper \cite{gdkcikk} I performed steady state 
simulations such that offsprings were generated on the left 
or right sides of parents $A \to ABB$, $A \to BBA$, $B \to BAA$
and $B \to AAB$.
The relaxation time in this case was very long, 
typically $t>10^6$ MCS that prohibited getting closer to the 
critical point than $\sigma=0.03$.
I performed these simulations in system sizes of 
$L=10^5$ sites and averaged over $100$ samples. 

\begin{figure}
\epsfxsize=70mm
\centerline{\epsffile{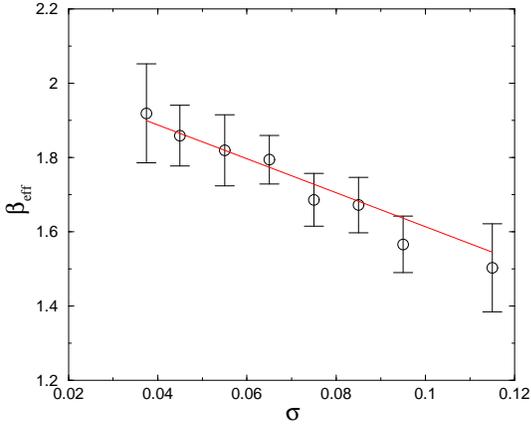}}
\vspace{2mm}
\caption{Effective order parameter exponent ($\beta$) results for
the asymmetric branching case.
\label{beta}
}
\end{figure}

As one can see in Fig.\ref{beta} the $\beta_{eff}$ 
extrapolates to $2.05(10)$ linearly
in agreement with what was deduced from the local mean-field 
approximation by Kwon {\it et al.} \cite{Kwon}. 
This value however, differs from the field theoretical 
predictions ($\beta=1$) \cite{Cardy-Tauber} obtained 
for the coarse grained, bosonic description of the 
particle system, which neglects hard-core interactions.

\begin{figure}
\epsfxsize=70mm
\centerline{\epsffile{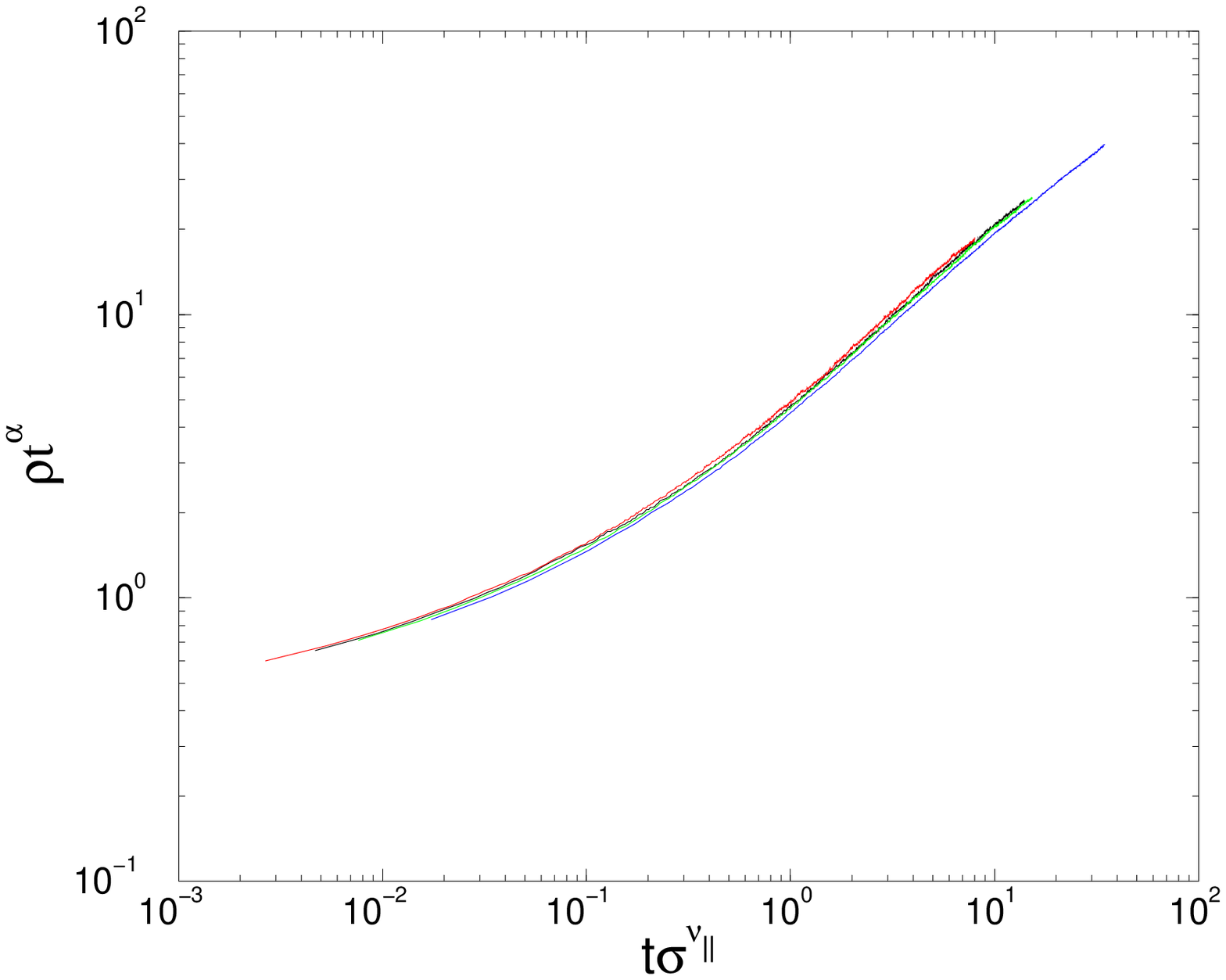}}
\vspace{2mm}
\caption{Data collapse for off-critical simulations of the
symmetric branching model according to the scaling form
(\ref{rhoscal}) with $\alpha^p=0.545$ and $\nu^p_{||}=3.66$.
The different curves correspond to $\sigma=0.03, 0.035, 0.04, 0.05$
and pairwise initial conditions.
\label{scol}
}
\end{figure}

I investigated the decay by starting from other different 
initial conditions as well. The steady state concentration 
and hence the exponent $\beta$ was found to be the 
same, independent of the initial density of particles.
In the case of random initial distribution of particles
the relaxation was found to be even longer reflecting 
the fact that initial alternating sequences 
slow down the evolution to the steady state but the
final concentration was the same.
Again by comparing the $Z$'s of Sec. \ref{sigmac}
with $\nu_{\perp}=2$ of \cite{Kwon} determined 
by static finitesize scaling for the asymmetric
branching process, we can predict for pairwise initial 
conditions  $\nu_{||}^p = Z \nu_{\perp} = 3.66(2)$ and
$\nu_{||}^r=8.0(4)$ for random initial conditions.
As in Sec. \ref{static} the standard scaling relation 
of the density is fulfilled for both initial cases
within error limits:
$\alpha^p\simeq 0.545 \simeq \beta / \nu_{||}^p = 2 / 3.66$
$\alpha^r\simeq 0.25 \simeq \beta / \nu_{||}^r = 8 / 2$
and the scaling law can be confirmed by data collapse
(see Fig. \ref{scol}).

In the case of dynamic branching Kwon {\it et al.} \cite{Kwon} 
claim a slight dependence (a few percent) of the static 
exponents on the diffusion rates of different species.
The relaxation time in this case is so huge
that it hindered achieving such a precision by simulations.

%=============================================================
\subsection{The off-critical case ($\sigma>0$) without exclusion}
%=============================================================

Finally I tested the field theoretical results in one dimension 
in such a way that I allowed particles of different types to 
exchange sites upon meeting. The steady state density
has been determined for different values of $\sigma$ around zero
in lattices with $L=2\times 10^4$ size. As one can see in 
Fig.\ref{rho-s} the middle curve (corresponding to this
case) approaches to $\sigma=0$ linearly indeed in agreement with
the $\beta=1$ field theoretical result. This can be verified
by considering $\beta_{eff}$ as well.

%=========================================================================
\section{Conclusions}
%=========================================================================

In this study I confirmed numerically the field theoretical
predictions for the two dimensional branching and annihilating 
random walk process of two species. The logarithmic corrections
to the power-law density decay function have been determined.

\begin{figure}
\epsfxsize=70mm
\centerline{\epsffile{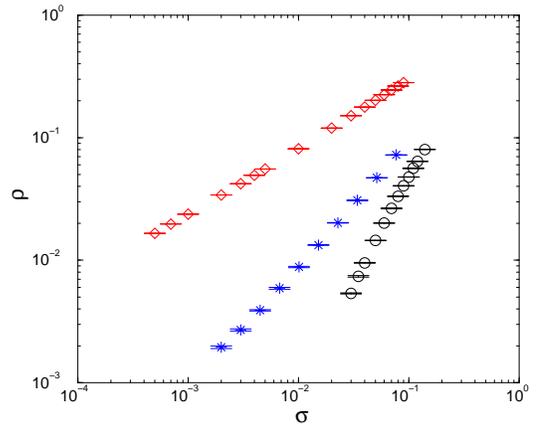}}
\vspace{2mm}
\caption{Steady state density in the one-dimenional
2-BARW-2 model. Circles correspond to the asymmetric 
branching with $\beta=2$, diamonds to the symmetric 
branching with $\beta=1/2$ and stars to the model 
without exclusion ($\beta=1$).
\label{rho-s}
}
\end{figure}

In one dimension I gave numerical evidence that the bosonic 
renormalization group field theoretical predictions 
cannot be applied in contrast to widespread beliefs
for systems with exclusion and the order parameter exponent
depends on the spatial symmetry of the offspring production process.
As Fig. \ref{rho-s} shows, the steady state density is much lower
if offsprings are created on the same side of the parent
against the case when they are separated by the parent resulting in
different scaling exponents $\beta=2$ versus $\beta=1/2$.
This particle exclusion effect results in new universality 
classes in one dimension as summarized in Table \ref{tab}
and the static exponents are insensitive to the initial
conditions.
In the case of static branching the precise numerical analysis 
of the order parameter data does not show dependence 
on the diffusion ratios either.

The density decay exponent starting from a random initial state 
($\alpha$) has been shown to agree with predictions of 
\cite{Krap-Red} but initial density-dependent logarithmic
corrections to scaling can be observed. 
The dynamical exponent $Z$ has been found to depend on
the initial conditions too. For pairwise initial 
conditions it is in agreement with our earlier 
study \cite{gdkcikk}. 
All these results for dynamical scaling deviate from 
those obtained by the continuum field theory and 
assumed in \cite{Kwon}.

This study gives further evidence to the claims of Kwon 
{\it et al.}\cite{Kwon} that in one dimension a series of new universality 
classes emerge in nonequilibrium particle systems exhibiting 
hardcore interactions and continuous phase transitions.
The relative simplicity of this model and the precision of the
critical exponents obtained may provide 
a good starting point to more elaborate analytical 
investigations of systems with hardcore exclusion.

\bigskip
\noindent
{\bf Acknowledgments}\\

The author would like to thank Z. R\'acz, S. Redner and U. T\"auber 
for stimulating discussions.
Support from Hungarian research fund OTKA (Nos. T-25286 and 
T-23552) and from B\'olyai (No. BO/00142/99)
is acknowledged.

\end{multicols}

\begin{table}
\begin{center}
\begin{tabular}{|l|r|r|r|r|r|}
process     & $\nu_{||}$ & $\nu_{\perp}$ & $Z$ & $\alpha$       & $\beta$ \\ 
\hline
nonblocking & 2         &      1   &  2  & 1/2            & 1       \\
\hline
symmetric & 2.0(1)/0.915(2) & 0.50(1) & 4.0(2)/1.82(2)*& 0.25(1)/0.55(1)*& 0.50(1) \\
\hline
asymmetric& 8.0(4)/3.66(2)  & 2.0(1) & 4.0(2)/1.82(2)*& 0.25(1)/0.55(1)*& 2.05(10)\\
\end{tabular}
\end{center}
\caption{Summary of critical exponents in one dimension.
The non-blocking data are quoted from [4].
%\cite{Cardy-Tauber}.
The predictions for $\nu_{\perp}$ of
the blocking models are cited from [18].
%\cite{Kwon}.
Data divided by "/" correspond to random vs. pairwise
initial condition cases.
%\cite{gdkcikk}.
Exponents denoted by * exhibit slight initial density dependence [17].
}
\label{tab}
\end{table}

\end{document}